\documentclass[structabstract]{aa}  

\usepackage{graphicx,psfig}
\usepackage{txfonts}

\begin{document}

\title{Ground-based detection of sodium in the transmission spectrum of exoplanet HD209458b}

\authorrunning{Snellen et al.}
\titlerunning{Ground-based detection of sodium in HD209458b}

\author{Snellen I.A.G., Albrecht S., de Mooij E.J.W. \and Le Poole R.S.}

\institute{Leiden Observatory, Leiden University, Postbus 9513, 2300 RA, Leiden, The Netherlands}   

\offprints{snellen@strw.leidenuniv.nl}

\date{}

\abstract{The first detection of an atmosphere around an extrasolar planet 
was presented by Charbonneau and collaborators in 2002. In the optical 
transmission spectrum of the transiting  exoplanet HD209458b,
 an absorption signal from sodium was measured at a level of 0.023$\pm$0.006\%, 
using the STIS spectrograph on the Hubble Space Telescope. Despite several 
attempts, so far only upper limits to the Na\,D 
absorption have been obtained using telescopes from the ground, and the 
HST result has yet to be confirmed.}
{The aims of this paper are to re-analyse data taken with the High Dispersion 
Spectrograph on the Subaru telescope, to correct for systematic effects
dominating the data quality, and to improve on previous results 
presented in the literature.}
{The data reduction process was altered in several places, most importantly 
allowing for small shifts in the wavelength solution. The relative depth of 
all lines in the spectra, including the two sodium D lines, are found to 
correlate strongly with the continuum count level in the spectra. These
variations are attributed to non-linearity effects in the CCDs.
After removal of this empirical relation the uncertainties in the line depths
are only a fraction above that expected from photon statistics.}
{The sodium absorption due to the planet's atmosphere is detected at 
$>$5$\sigma$, at a level of 0.056$\pm$0.007\% (2$\times$3.0\AA$ $ band), 
0.070$\pm$0.011\% (2$\times$1.5 \AA$ $ band), and 0.135$\pm$0.017\% (2$\times$0.75\AA$ $ band). There is no evidence that the planetary absorption signal is shifted with respect to the stellar absorption, as recently claimed for HD189733b.} 
{The STIS/HST measurements are confirmed. The measurements of the Na\,D absorption in the two most narrow bands indicate that some signal is being resolved.
Due to variations in the instrumental resolution and 
intrinsic variations in the stellar lines due to the Rossiter-McLauglin effect, it will be challenging to probe the planetary absorption on spectral scales smaller than the stellar absorption using conventional transmission spectroscopy.}

\keywords{techniques: spectroscopic  -- stars: atmosphere  -- stars: planetary systems}

   \maketitle

\section{Introduction}

Transiting extrasolar planets are of great scientific value. 
While the radial velocity method continues to be very successful in finding
planets and characterising their orbits, only transits can currently
reveal the properties of the planets themselves. In addition to the basic
planetary parameters that can be determined, such as planet mass, size, and 
average density, the atmospheres of transiting planets can be probed through
either secondary eclipse observations (e.g. Charbonneau et al. 2005; Deming et al. 2005; Knutson et al. 2007) or atmospheric transmission 
spectroscopy, the subject of this paper. 
In transmission spectroscopy, the depth of a planet transit 
is measured as function of wavelength. It is expected that at certain 
wavelengths, a transit will be slightly deeper due to absorption
in the planet's atmosphere. In the optical transmission spectrum of hot 
Jupiters, 
the strongest of these absorption features was predicted to come from the 
sodium D lines at 5889 and 5896 \AA$ $ (Brown 2001; Seager \& Sasselov 2000). 
Indeed, Charbonneau et al. (2002) detected Na\,D absorption 
in the transmission spectrum of the transiting exoplanet HD209458b, at a level
of 0.023$\pm$0.006\% in a 12\AA$ $ wide band, using the STIS spectrograph 
on the Hubble Space Telescope (HST).
This constitutes the first detection of an atmosphere around an extrasolar 
planet. Subsequently, strong absorption features have been detected
in HD209458b from hydrogen, oxygen, and carbon at a level of 5-15\%, 
thought to be caused by an evaporating exosphere (Vidal-Madjar 
et al. 2003; 2004). Recently, hot hydrogen has been detected by
Ballester et al. (2007), also using HST data. 
A claim by Tinetti et al. (2007) of the 
detection of water vapour from a comparison of transit depths at several 
wavelengths 
in the infrared, as measured with Spitzer, has been disputed by Ehrenreich 
et al. (2007). This, while Swain et al. (2008) identify both water and methane
from NICMOS/HST data.

Despite several attempts from ground-based observatories, no confirmation has 
yet been obtained of the Na\,D 
planetary absorption feature in the transmission spectrum of HD209458b.
In general, ground-based transmission spectroscopy has not been a great success.
Typically, upper limits to a Na\,D absorption signal of 0.1$-$1\% have
been reached (Moutou et al. 2001; Snellen 2004; Narita et al. 2005),
implying that systematic effects dominate the error budgets. 
A modern Echelle spectrograph on a 8$-$10 m. class telescope can provide
spectra from the brightest transiting exoplanet systems with signal-to-noise ratios,
SNR$>$100, within a few minutes of exposure time.
Integrating over a few Angstrom and over the duration of a transit, this would 
mean that photon noise statistics should allow detections down to a few times 
10$^{-4}$. Although the HST detection of the Na\,D absorption feature is only
just at this level, it is expected that its width is only a fraction of 
the 12\AA$ $ passband used by Charbonneau et al. (2002), as recently shown by
re-analysis of the STIS data (Sing et al. 2008). 
This means that the Na absorption
within a 2-3\AA$ $ band should be at the $\sim10^{-3}$ level.

In this paper we re-analyse a data-set from the High Dispersion Spectrograph on the Subaru telescope, that covers one transit of HD209458b. 
The aim is to identify and correct for possible systematic effects,
and to improve on the results previously presented by Narita et al. 
(2005; NAR05). Their analysis resulted in 
spectra with an SNR of a few hundred in the stellar continuum. However, near 
strong absorption lines, such as the Na\,D doublet, clear coherent structures
were visible (Figure 2 in NAR05), well in excess of the photon shot noise.
While NAR05 argued that the positions of the spectral lines
are stable to within 0.01\AA, spectral shifts at only a fraction of this level 
(e.g. due to slit-centering variations) could cause these effects.
This encouraged us to analyse this data again.
In Section 2, the observations, data reduction and analysis are described. 
The result on the Na\,D absorption are presented and discussed in Section 3,
together with a comparison with the STIS/HST results, and with a
recent detection of Na\,D absorption in exoplanet HD189733b (Redfield et al.
2008)

\section{Observations, data reduction, and analysis}
\begin{figure}
\psfig{figure=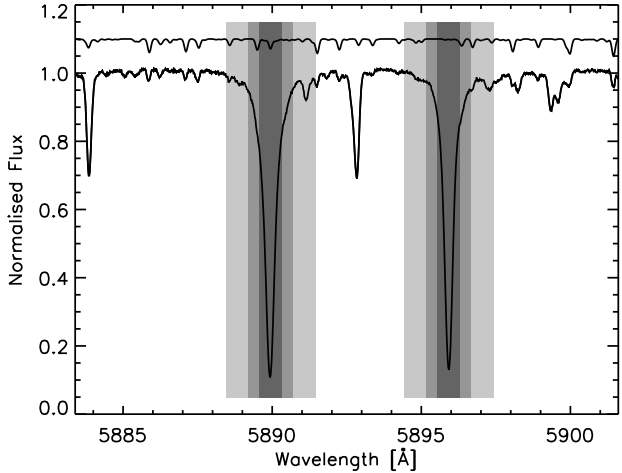,width=0.5\textwidth}
\caption{\label{spectrum} Spectrum of HD209458 around the Na\,D doublet.
 The shaded areas indicate the 
narrow, medium, and wide passbands as used in our analysis. In all cases, 
the comparison bands are located directly adjacent to the central band,
and have the same width. The upper line shows a synthetic telluric spectrum 
constructed from the line list of Lundstrom et al. (1991). Note that the 
telluric sodium absorption can show strong seasonable variability.
}
\end{figure}

HD209458 was observed on the night of October 24, 2002, using the High 
Dispersion Spectrograph (HDS; Noguchi et al. 2002) on the Subaru telescope. 
We obtained the data using the SMOKA archive system (Baba et al. 2002).
The observations have been described in Winn et al. (2004) and 
Narita et al.(2005; NAR05). Thirty-two spectra were taken in {\sl Yb mode} 
(without the iodine cell), of which the last thirty were made with an exposure time 
of 500 seconds. The entrance slit was 4$''$ long and 0.8$''$ wide, oriented
with a constant position angle, resulting in a spectral resolution
of R$\sim$45\,000 with 0.9 km s$^{-1}$ per pixel. We concentrated on the 
data from the red CCD, which contains 21 orders of 4100 pixels covering 
5500\AA$<\lambda<$6800\AA. Twelve of the thirty spectra fall outside 
the transit (nine before ingress and three after egress), and eighteen 
during the transit. The uncertainty in the transit timing is neglegible.

For the initial data reduction we followed the procedure of NAR05. 
First the frames were processed using the IRAF software package, including
the extraction of the one-dimensional spectra.
These spectra have signal-to-noise ratios that vary between 300 and 450
per pixel in the continuum. Subsequent analyses were
conducted using custom-built procedures in the Interactive Data Language (IDL). 
Winn et al. (2004) and NAR05 describe a good method to correct for 
time-dependent variations of the instrumental blaze function 
( possibly caused by flexure of the spectrograph) using the 
adjacent orders, which we also use here. We do not apply a global 
wavelength solution to the spectra. Only for the analysis of the strength 
of telluric absorption features (see below) did we apply the wavelength 
solution to the order containing the Na\,D doublet. 
The Na\,D spectral surroundings of HD209458 are 
shown in Figure \ref{spectrum}.

\begin{figure}
\psfig{figure=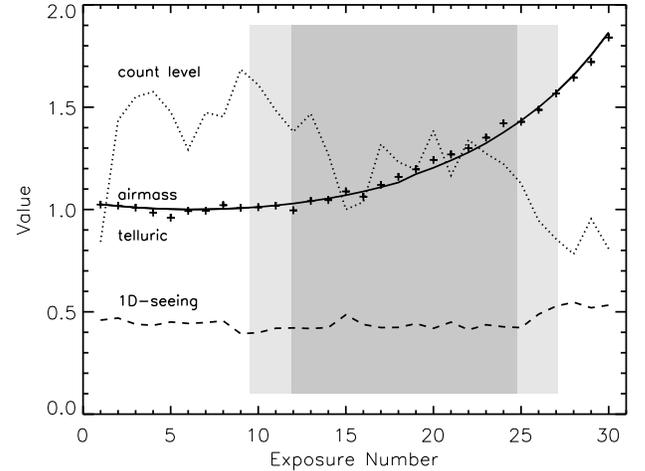,width=0.5\textwidth}
\caption{Variations of observing conditions over the 30 spectra. 
The solid line 
indicates the airmass, and the dotted line shows the normalised variation
in continuum count level. The dashed line indicates the seeing (in $\sigma$) as 
measured from the one-dimensional profile of the star along the slit. The 
crosses show the relative telluric line strength as measured using strong lines
in the red part of the spectrum. It follows the airmass well, 
implying that the conditions were excellent. The grey band indicates the 
timing of the planetary transit and periods of ingress and egress.\label{circum}}
\end{figure}

The variations in observing conditions during the night are shown in Fig. \ref{circum}. The solid line indicates the variation in airmass, and the dashed line 
 the variation in the 1-dimensional stellar profile along the slit 
(a measure of the seeing). The dotted line shows the normalised 
variation in continuum count level in the spectra, indicating that it varies
up to a factor of two from spectrum to spectrum.

\begin{figure}
\psfig{figure=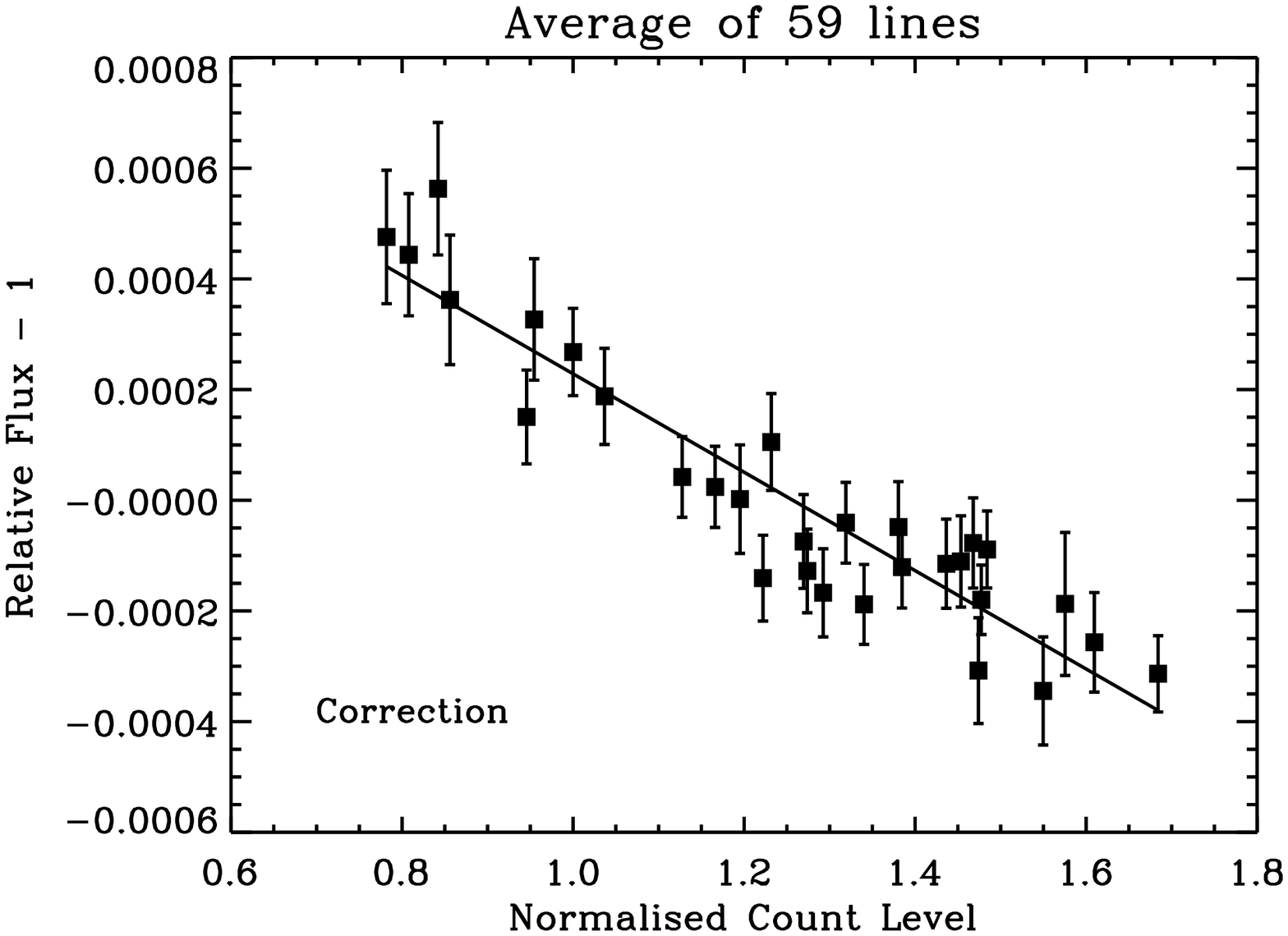,width=0.5\textwidth}
\psfig{figure=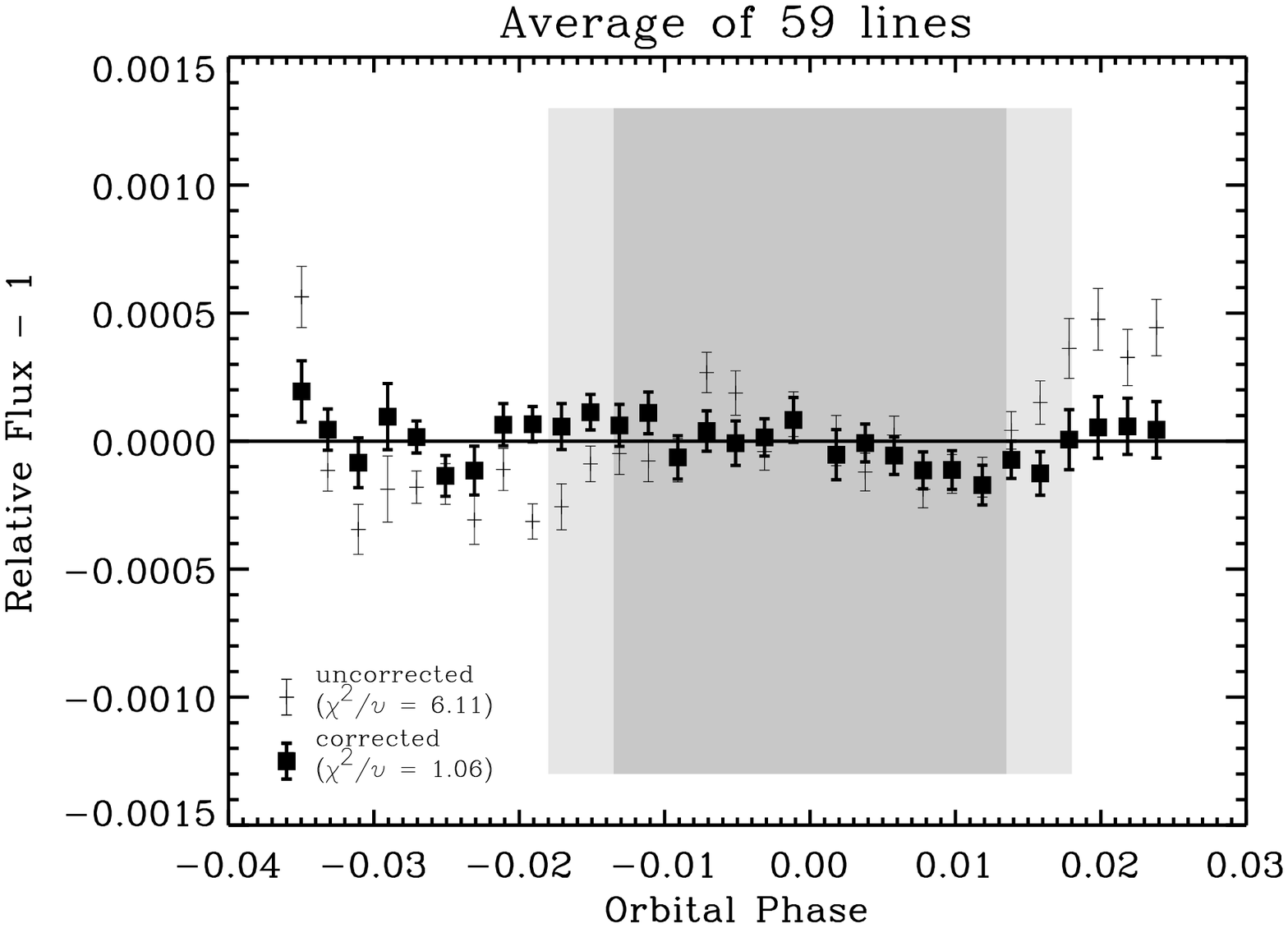,width=0.5\textwidth}
\caption{\label{other} The top panel shows the normalized count level of the 
 continuum in the spectra versus the measured weighted mean-strength of 
59 deep stellar absorption lines. A strong inverse correlation is present,
with the more flux in the spectra, the deeper the absorption lines.
We believe that this is due to a non-linearity effect in the CCD. The solid line indicates the linear least squares fit to the correlation, used to correct the measured line strengths. The lower panel shows the same average strength of  these
59 absorption lines as function of orbital phase, where the grey band indicates the planetary transit. Those data points corrected for the correlation with count level (thick squares) have a reduced chi-squared of 1.06, while the uncorrected data exhibit a reduced chi-squared of 6.11.}

\end{figure}

We subsequently selected 59 strong stellar lines and the Na\,D doublet to 
perform our analysis. First, for each line, the line center, $\lambda_0$, was 
determined through Gaussian fitting. Subsequently, the total flux was 
integrated within 
a spectral band, $\Delta \lambda$, and divided by the average of two
equally wide bands to the left and right of the line.

\begin{equation}
\left.
\begin{array}{r}
F_{\rm{mid}}  = \int_{\lambda_0-\frac{1}{2}\Delta \lambda}^{\lambda_0+\frac{1}{2}\Delta \lambda}
F(\lambda)d\lambda  \ \ \ \\
F_{\rm{left}} = \int_{\lambda_0-\frac{3}{2}\Delta \lambda}^{\lambda_0-\frac{1}{2}\Delta \lambda}
F(\lambda)d\lambda  \ \ \ \\

\ \ \ \ \ F_{\rm{right}} = \int_{\lambda_0+\frac{1}{2}\Delta \lambda}^{\lambda_0+\frac{3}{2}\Delta \lambda}
F(\lambda)d\lambda \ \ \ 
\end{array} \right\}  \ \ \ \ \
F_{\rm{line}} = \frac{2 F_{\rm{mid}}}{F_{\rm{left}}+F_{\rm{right}}}
\end{equation} 
Subsequently, the relative flux in each line is divided by its average 
value on the night. Three spectral widths were used of 42, 84 and 166 pixels 
wide, corresponding to $\Delta \lambda$ = 0.75 \AA$ $, 1.5 \AA$ $, and 3.0 \AA$, $ respectively, at the Na\,D doublet.

\subsection{Systematic effects}

The variations from spectrum to spectrum, away from spectral lines, are, as 
expected, dominated by photon noise. However, at and around strong 
 lines, systematic effects play a role. This is in addition to the
fact that the signal-to-noise ratios in the cores of deep lines are 
significantly lower. 
Most importantly, it was noticed that the observed depths of the Na\,D lines 
are 
a clear function of the overall count level in the spectra. 
The same effect is visible
 in the average of the 59 reference lines. In the top panel of Fig. 
\ref{other}, for each spectrum, the normalised continuum level is plotted 
against the weighted (by line-strength) mean depth of the lines.
It shows that the higher the count
 levels in the spectra, the deeper the lines. Firstly, we convinced ourselves 
that this is not due to scattered light residuals. It would need about 300 
counts per spectral bin to cause this effect, two orders of magnitude above the 
uncertainties in the scattered light removal. Instead, several other possible causes are identified, 
but we believe that non-linearity of the CCD is the most likely cause. Other 
potential systematic effects are caused by variations in the seeing and/or 
spectral resolution, and intrinsic variations of the stellar spectral line 
due to the Rossiter-McLaughlin effect (Rossiter 1924). 
The latter two effects result in strong 
variations across deep spectral lines, making atmospheric transmission 
spectroscopy very challenging at spectral scales shorter than the width of the 
stellar lines. 

\begin{itemize}
\item[1] {\bf Non-linearity of the CCD:}
Most CCD arrays show low level non-linearity effects, meaning that 
the conversion factor between electrons and data numbers (ADU) is not 
constant, but is slowly varying as function of the count-level in a pixel.
At levels below the half full well, the conversion factor often 
decreases towards higher count levels. By assuming a  completely linear
CCD with constant gain, high count levels are over-estimated, and low count 
levels (such as in the center of strong absorption lines) are relatively 
under-estimated.
This will make stellar lines appear deeper in those spectra that are better
exposed, just as is seen in Fig.~\ref{other}. 
Note that the first of the thirty frames was exposed for only 300 sec instead
of 500 sec. The lines in this spectrum also follow the relation between line
depth and count level. This is in agreement with a non-linearity effect.
Unfortunately, to our knowledge, the non-linearity has not been measured for
the HDS CCDs. We also found that subtle changes in the order profiles from 
exposure to exposure makes it impossible to estimate the non-linearity for 
the CCDs from the flat field exposures.
Our tests with a model for the 
non-linearity of the CCD indicates that a decrease in gain of the order of 
2-3\% towards 10\,000 ADU would be sufficient to explain the effect seen 
in Fig. \ref{other}.  By adjusting the raw ADU count levels to
$F_{\rm{cor}}=(1.0 - 0.03 \times F_{\rm{raw}}/ 10^{4})F_{\rm{raw}}$ before
the reduction process, the effect is largely removed.
 Note that this is within the range of non-linearity 
effects seen in other EEV42-80 type CCDs \footnote{e.g. the CCDs of the Wide Field Camera on the Isaac Newton Telescope: http://www.ast.cam.ac.uk/$\sim$wfcsur/technical/foibles/}.

\item[2] {\bf Variations in spectral resolution and seeing.}
Variations in the spectral resolution will also result in variations in the 
depth of stellar lines. The equivalent width of a line should not 
vary with spectral resolution, meaning that the difference between two spectra 
with different spectral resolutions should always integrate to zero over 
a line. However, the ratio of the two spectra can show strong variations 
across a line, e.g. because in one of the spectra the line 
appears much deeper and sharper than in the other spectrum.
The latter does not integrate to zero, in 
particular for very strong lines such as the Na\,D doublet. 
Since the HDS is a 
slit-spectrograph, the seeing influences the exact spectral resolution,
and therefore this effect could be present. 
If the seeing is significantly larger than the slit width, the star light is 
homogeneously distributed across the slit. If the seeing is similar to the 
slit width, or smaller, the light distribution will be peaked, resulting in a 
higher spectral resolution. This is also the reason why in general 
slit-spectrographs do not offer the highest stability for radial velocity 
measurements.

\item[3] {\bf Intrinsic variations of the stellar lines}
During the transit, the intrinsic shapes of the stellar lines also change due
to the Rossiter-McLauglin effect (Rossiter 1924). 
When the planet disk crosses the star, it first blocks off part of the stellar 
surface that due to the star's rotation is moving towards us, and later blocks 
part of the  stellar surface that is moving away from us. This 
not only results in a radial velocity abberation, but also alters the overall 
spectral line shape. As above, the equivalent widths of the 
lines do not change, but the ratio of spectra taken during and outside the transit 
will show features that do not integrate down to zero.  

\end{itemize}

Changes in the spectral resolution are unavoidable in slit-spectrographs such
as the HDS, but should be largely absent in fiber-fed spectrographs. However, 
the intrinsic variations due to the Rossiter-McLauglin effect will be present 
in all transmission spectroscopy observations. To avoid the consequences of 
line shape changes the flux first should be integrated over the total width of 
the line before 
it can be compared between spectra, meaning that $\Delta \lambda$ in 
equation 1 should be large enough. We therefore do our analysis with a minimum 
width of 0.75 \AA.

Most likely, the non-linearity of the pixels in the HDS CCD is causing the 
effect seen in the top panel of Fig. \ref{other}. Since it was not possible
to directly measure the non-linearity of the CCD, we corrected for it 
in an empirical way, by performing a least-squares fit to the correlation
between line depth and count level.
For the weighted mean of the 59 strong comparison lines, the reduced 
chi-squared drops from $\chi^2/\nu$=6.11 to $\chi^2/\nu$=1.05 after 
this correction. The bottom panel of Fig. \ref{other} shows the 
remaining residuals. Although there is some correlated 
variation still visible, it averages out over the transit.
Since we integrate over the total extent of the lines, we do not 
expect variations in the seeing to affect our results. Nevertheless, since
the seeing influences how much flux of the star enters the slit, the seeing
is anti-correlated with the count level in the spectra. Therefore,
a correlation between the seeing and the average line strength is also 
present. However, the resulting reduced chi-squared is $\chi^2/\nu$=2.90,
significantly higher than that resulting from the continuum count level,
indicating that changes in seeing are not the underlying cause of the line 
variations.

\subsection{The Na\,D lines and telluric contamination}

In a similar manner as for the weighted mean of the reference
lines, the correlation between line depth and continuum count level was 
removed for the Na\,D doublet. We subsequently determined the depth of 
the transit for the three Na\,D passbands separately. A least-squares
fit was performed on the data using as a model a scaled version of the 
HST light curve as presented by Brown et al. (2001).  
The best fitting model is subsequently removed from the data, revealing
low-level variations due to changing telluric contaminations most evident
at the end of the night. This telluric residual was found to scale with
the airmass and with the strength of some strong telluric lines in the 
red part of the spectrum (as shown in Fig. \ref{circum}). A linear fit
was performed between the average strength of the strong telluric lines and 
the Na\,D residuals to remove the telluric contribution.
Note that the fitting of the continuum count level, transit signal, and 
telluric contamination was performed in a iterative way, but the solutions
did not significantly change after the first round. 

In addition, a consistency check was performed to see whether our 
telluric line contamination removal was reasonable. Many telluric lines are present 
around the Na\,D doublet, in both the line and reference bands. 
We used the telluric line list of Lundstrom (1991) to construct a synthetic 
telluric spectrum. We then constructed a reference spectrum from the average
of all exposures which was subsequently removed from the 30 frames.
Although the absolute telluric contamination is now lost, all information
on the change in telluric contamination becomes clearly visible in this way. 
The synthetic telluric spectrum was then fitted to these frames, and 
the change in telluric contribution in the various spectral bands determined.
This gives very similar results as the method described above. 
We use the former method in our final analysis since the contribution of 
telluric sodium relative to that of water and oxygen undergoes seasonal 
variations, but within one night is expected to vary following the 
other telluric lines.
Note that NAR05 used the 
spectrum of the rapidly rotating B5 star HD42545 to remove the telluric 
contamination around the Na\,D doublet, however the sodium absoption
towards this star is dominated by interstellar contributions (although 
this is unlikely to have influenced their results).

\section{Results and discussion}

\begin{figure}
\psfig{figure=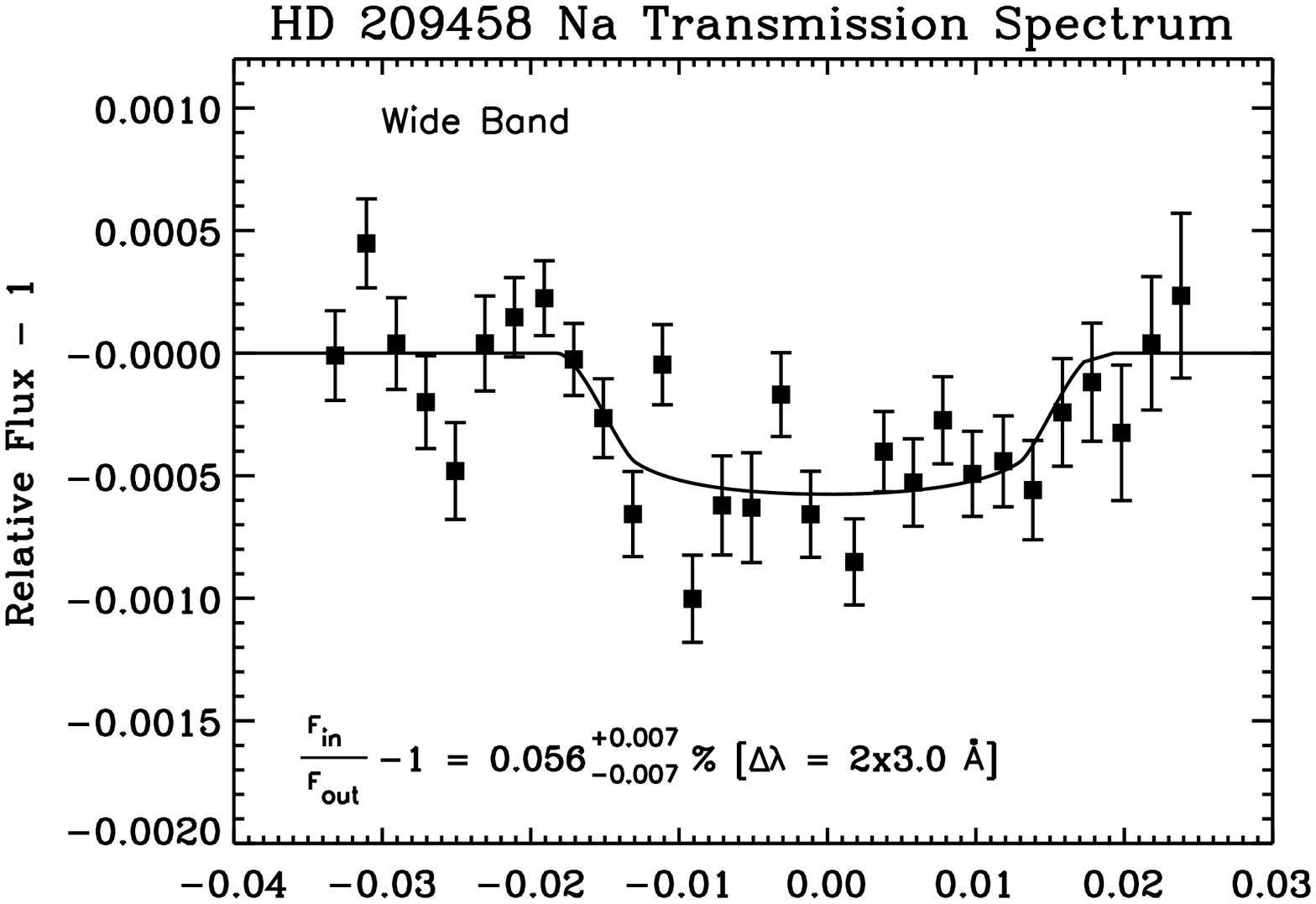,width=0.5\textwidth}
\vspace{-1cm}
\psfig{figure=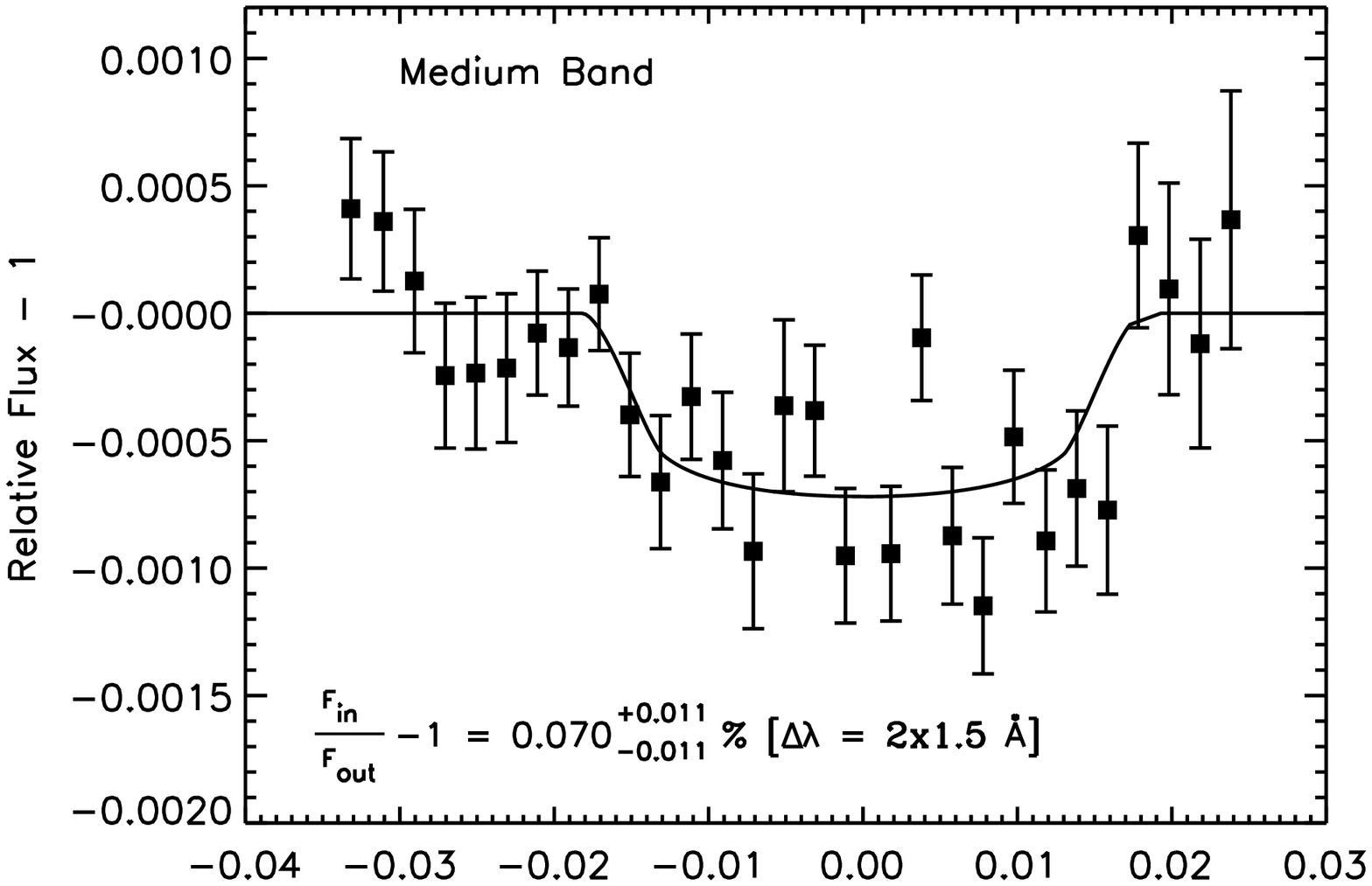,width=0.5\textwidth}
\vspace{-1cm}
\psfig{figure=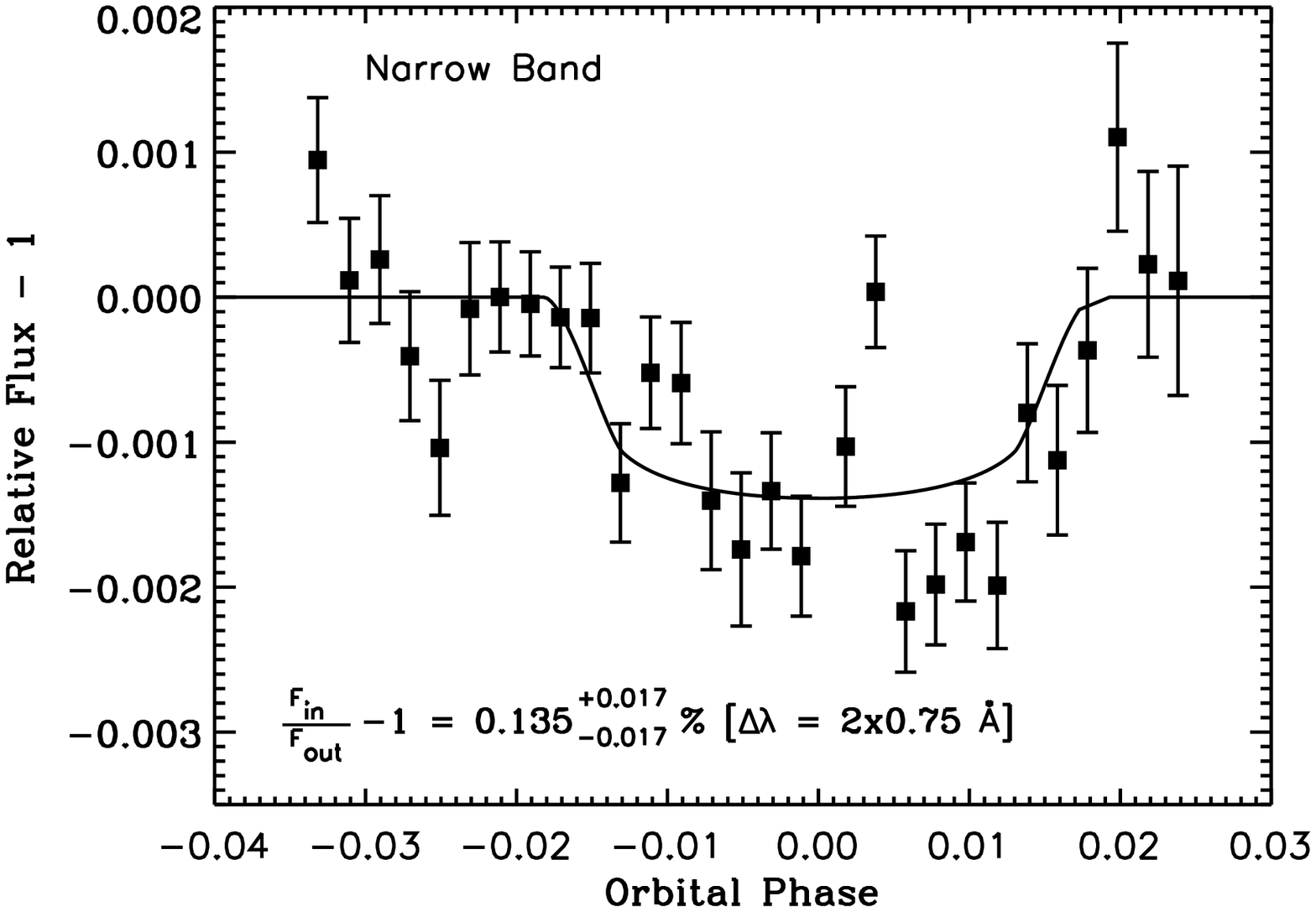,width=0.5\textwidth}

\caption{Transit photometry averaged in two bands centered on the 
Na\,D doublet, with spectral widths of 3$\AA$ (top panel), 1.5 $\AA$ (middle
panel) and 0.75 $\AA$ (bottom panel).  The data have been empirically corrected for the dependence of the depth of stellar lines on the count level in the spectra, as discussed in detail in section 2.1. \label{res1}}
\end{figure}

\begin{figure}
\psfig{figure=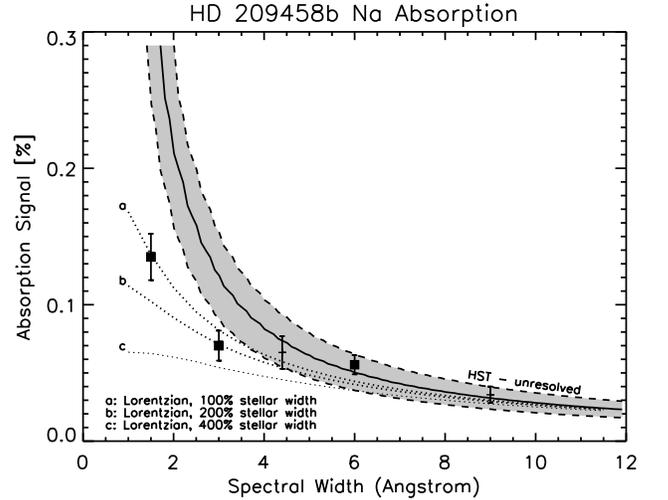,width=0.5\textwidth}
\caption{\label{widths} The three measurements of the Na\,D absorption in
the transmission spectrum of HD209458b in passbands of 2$\times$0.75\AA,  2$\times$1.5\AA, and  2$\times$3.0\AA. The solid line and grey band indicate the HST 
measurement by Charbonneau et al. (2002) assuming it is completely unresolved and within our passbands. The three dotted lines, indicated by
 $a$, $b$, $c$,  are the Na\,D absorption levels as a function of passband  
expected from the HST measurements, assuming that the atmospheric absorption 
feature is Lorentzian shaped with  a width equal to, 2$\times$, and 4$\times$ the 
stellar width of the Na\,D doublet. The two crosses at 4.4 and 9 $\AA$ $ $ are the results from the re-analysed STIS/HST data by Sing et al. (2008a).}
\end{figure}

We measured the depth of the Na\,D features in the transmission spectrum
within three spectral passbands centered on the two stellar lines,
with widths of 0.75\AA, 1.5\AA, and 3.0\AA. The results are presented in 
Fig. \ref{res1}.
The sodium absorption due to the planet's atmosphere is detected at 
$>$5$\sigma$, at a level of 0.056$\pm$0.007\% (2$\times$3\AA$ $ band), 
0.070$\pm$0.011\% (2$\times$1.5\AA$ $ band), and 0.135$\pm$0.017\% (2$\times$0.75\AA$ $ band).  The quoted uncertainties are 1$\sigma$ error intervals as
determined from the SNR in the spectra using chi-square analysis.
The resulting reduced chi-squares values, $\chi^2/\nu$, are 47/28=1.63, 31/28=1.10, and 45/28=1.75 respectively, and 
indicate that the residual noise levels are 10$-$30\% higher than expected 
from Poisson statistics. A way to take this residual noise into account in the
error budget is to scale the error bars up such that $\chi^2/\nu$=1.
This would increase the uncertainties by 10$-$30\%, resulting in 
conservative estimates of the significance of the sodium detection of 
$\sim$6$\sigma$ in each individual passband.

A crucial part of our data analysis is the empirical correction for the 
correlation of line depth with the continuum count level in the spectra,
attributed to non-linearity effects in the CCD.
To assess the robustness of our result we performed a slightly different
empirical correction by directly correlating the weighted mean-strength 
of the 59 reference lines with that of the Na\,D lines. 
Depending on the passband, this alternative analysis results in a transit
depth 20$-$30\% lower than determined above, also with 
 20-30\% higher $\chi^2/\nu$ values. Since this would still be $\sim$5$\sigma$ 
detections, it further supports the detection. 
Since the
first method gives significantly less noisy results, we believe that
method it is more reliable. 

In our analysis we do not take into account the variation
in radial velocity of the planet (and its atmospheric absorption) relative to that
of the star. During the 
transit, the radial velocity of the planet  varies from about $-$14 to +14 
km sec$^{-1}$. For a Lorentzian shaped planetary absorption profile with a width
comparable to that of the star (see below), the strength of the absorption
signal will have been underestimated by $\sim$4\% in the narrowest passband.
Therefore this is just a minor effect.

\subsection{Comparison with the STIS/HST results}

Our results are in excellent agreement with the STIS/HST results presented by
Charbonneau et al. (2002) and Sing et al. (2008a).
Fig. \ref{widths} shows our three measurements of the planetary Na\,D 
absorption as a function of the spectral passband.
The solid line and grey band shows the STIS/HST result and its uncertainty
corrected to our passbands, assuming that the planetary absorption is 
unresolved and completely within them.
If there were no {\sl stellar} Na\,D lines, then the expected absorption strength as a function of passband would simply scale as $\Delta \lambda^{-1}$.
However, since most absorption occurs near the cores of the stellar 
Na\,D doublet, where there is already little flux present, 
the relative contribution of the absorbed part of the spectrum varies more 
steeply than as 1/$\Delta \lambda$.
Our measurement in the 
widest passband is perfectly consitent with an unresolved HST absorption,
but some absorption appears to be missing in the two smallest passbands,
about 40\% in the 2$\times$1.5\AA$ $ band and about 60\% in the 
2$\times$0.75\AA$ $ passband, indicating that the planetary absorption is 
partially being resolved out. 
We modelled this by measuring the absorbed flux as function of passband 
for a planetary Lorentzian line profile with a width 
equal, 2$\times$ and 4$\times$ times the width of the stellar Na\,D absorption,
normalised to the STIS/HST value.
These simulations are shown as the dotted lines {\sl a, b, c} in Fig. 
\ref{widths}, and indicate that 
planetary absorption is likely 
to be about as broad (within a factor of 2) as the stellar absorption.
Also presented in Fig. 5 are two measurements from the re-analysis of 
the HST/STIS data (Sing et al., 2008a;b), within a band pass of 4.4 and 
9 $\AA$.
The analysis of Sing et al. is fully consistent with our results, 
but indicates that the Na line profiles may not be Lorentzian on wide scales. 
Instead they present narrow cores extending over a broader plateau.
The observations presented here are not sensitive to absorption on these large 
spectral scales.

\subsection{Comparison with the Na\,D absorption in HD189733}

Recently, Redfield et al. (2008) measured Na\,D absorption in the 
transmission spectrum of the other known bright transiting hot Jupiter
HD189733b, at a level of 0.067$\pm0.021$\% in a passband of 12\AA.
Note that their in-transit data originate from short observations of eleven 
different transit events, but in total reach a similar SNR as the data 
presented in this paper.
The detection of 
Redfield et al. is about a factor of 3 higher than that measured for HD209458b.
From their figure 1 we conclude that they have derived the strength of the 
Na\,D feature by integrating $\int_{\Delta \lambda}F_{in}/F_{out}$ instead of 
integrating $\int_{\Delta \lambda}F_{in}/\int_{\Delta \lambda}F_{out}$. This means
that their result cannot be directly compared to the results of HD209458b,
since it puts much higher weights ($>$10$\times$) 
on the pixels in the center of the stellar absorption lines where most planetary absorption 
is seen. 
The number of 
   absorbed photons for a $\sim$1\% planetary absorption at the stellar line 
center would be equal to the number of absorbed photons for a $\sim$0.1\% planetary absorption at the stellar continuum.
By multiplying the data points from 
their figure 1 by the stellar flux at each wavelength, we estimate that 
the measured Na\,D absorption from HD189733b is only a factor
$\sim$2 above that measured for HD209458b. This remaining difference  
however does not mean that the 
planets have a different Na atmosphere. Physically, these
levels of absorption correspond to a variation in the apparent planetary radius
of 0.8\% ($\sim$750 km) for HD209458b, and 1.0\% ($\sim$780 km) for 
HD189733b. Hence the results for both
planets are actually very similar. 

Remarkably, Redfield et al. (2008) find a significant blueshift of their 
planetary absorption signal, of the order of $\sim$38 km sec$^{-1}$ 
(corresponding to $\sim$0.75\AA). 
No such shift of the planetary absorption is present in 
the transmission spectrum of HD209458b. If this velocity shift is real,
it is rather puzzling how it could be produced, since it is about
an order of magnitude higher than the expected sound speed in the 
upper layer of the planet's atmosphere (Brown 2001). We note that the
data analysis of Redfield et al. is affected by the intrinsic variations
in stellar line shapes due to the Rossiter-McLaughlin effect, due 
to the use of the  $\int_{\Delta \lambda}F_{in}/F_{out}$ integral, which
does not have to integrate down to zero, even if there is no additional 
planetary absorption (see section 2). If the spectra are not evenly 
distributed over the 
transit, this can also lead to spurious velocity offsets, 
although it is difficult to see how these could become larger than the 
$v$sin$i$ of the star. 

\section{Conclusions}

We present the first ground-based detection of the Na\,D absorption feature in 
the transmission spectrum of the extrasolar planet HD209458b, 
fully consistent with the HST measurements by Charbonneau et al. (2002). 
The absorption is measured at a level of 0.135$\pm$0.017\%, 0.070$\pm$0.011\%,
and 0.056$\pm$0.007\% in three passbands of 2$\times$0.75\AA, 2$\times$1.5\AA, and 2$\times$3.0\AA$ $ wide,
indicating that the absorption is partially resolved out in the two smallest bands.
Crucial in our analysis was the removal of an empirical correlation between 
line depth  and the continuum count level in the spectra, likely to be caused
by non-linearity of the CCD. We show that due to either 
changes in the spectral resolution, or intrinsic changes in the stellar line 
profiles during the transit due to the Rossiter-McLaughlin effect, 
it is crucial 
to integrate the atmospheric absorption over a wide-enough passband 
to avoid spurious effects.

\begin{acknowledgements}
We thank the anonymous referee for his or her insightful comments.
Based on data collected at the Subaru Telescope and obtained from the SMOKA, which is operated by the Astronomy Data Center, National Astronomical Observatory of Japan.\end{acknowledgements}

\end{document}